\documentclass[fleqn,12pt,twoside]{article}
\usepackage[headings]{espcrc1}

\readRCS
$Id: espcrc1.tex,v 1.2 2004/02/24 11:22:11 spepping Exp $
\usepackage{graphicx}
\usepackage[figuresright]{rotating}

\newcommand{\AmS}{{\protect\the\textfont2
  A\kern-.1667em\lower.5ex\hbox{M}\kern-.125emS}}

\title{Scaling of anisotropy flows in intermediate energy heavy ion collisions
}

\author{Y. G. Ma\thanks{E-mail: ygma@sinap.ac.cn},
 T. Z. Yan, X. Z. Cai, J. G. Chen, D. Q. Fang, W. Guo, G. H. Liu, C. W. Ma, E. J. Ma,  W. Q.
Shen,  Y. Shi,  Q. M. Su, W. D. Tian, H. W. Wang, K. Wang
\address[SINAP]{Shanghai Institute of Applied Physics, Chinese
Academy of Sciences, P. O. Box 800-204, Shanghai 201800, China}%
       }

\runtitle{Scaling of anisotropy flows ...} \runauthor{Y. G. Ma et
al.}

\begin{document}

% typeset front matter
\maketitle

\begin{abstract}
Anisotropic flows ($v_1$, $v_2$ and $v_4$)  of  light nuclear
clusters are studied by a nucleonic transport model in
intermediate energy heavy ion collisions. The number-of-nucleon
scalings of the directed flow ($v_1$) and elliptic flow ($v_2$)
are demonstrated for light nuclear clusters. Moreover, the ratios
of $v_4/v_2^2$ of nuclear clusters show a constant value of 1/2
regardless of the transverse momentum. The above phenomena can be
understood by the coalescence mechanism in nucleonic level and are
worthy to be explored in experiments.
\end{abstract}

\section{Introduction}

Anisotropic flow can reflect the initial state of the heavy ion
reaction and evolution dynamics. In addition, the information of
the nuclear equation of state (EOS) could be also learned from the
flow studies
\cite{Olli,Voloshin,Sorge,Danile,Zhang,Shu,Kolb,Zheng,Gale,yg_prc,INDRA}.
Many studies of  the 1-th and 2-nd anisotropic flows, namely the
directed flow and elliptic flow revealed much rich physics in
heavy ion collision (HIC) dynamics. Recently, it was demonstrated
that the number of constituent-quark (NCQ) scaling exhibits from
the measurement of transverse momentum dependence of the elliptic
flow for different mesons and baryons in ultra-relativistic Au +
Au collision experiments at the Relativistic Heavy Ion Collider
(RHIC) in Brookhaven National Laboratory  \cite{J.Adams}, it
indicates that the partonic degree of the freedom  plays a
dominant role in the formation of the dense matter during early
stage of collisions. The coalescence or recombination models of
the constituent quarks  have been proposed to interpret the
NCQ-scaling of hadrons at RHIC
\cite{Ko,Molnari,Duke,Hwa,Chen,Ma-UCLA}. On the other hand,  the
coalescence mechanism has been also used to explain the formation
of light fragments  and their spectra of kinetic energy or
momentum  in intermediate energy heavy ion collisions some times
ago \cite{Awes,Mekjian,Sato,Llope,Hagel}. In addition, mass
dependence of the directed flow was experimentally investigated in
a few studies \cite{Huang,Kunde}. However, systematic theoretical
studies on the anisotropic  flow of different nuclear fragments in
intermediate energy domain in terms of the coalescence mechanism
is still rare.

In this paper,  we will investigate the  flow scaling of light
nuclear fragments in a framework of a nucleonic transport model,
namely isospin-dependent quantum molecular dynamics (IDQMD) at
intermediate energies. A naive coalescence mechanism in nucleonic
level was applied to analyze the fragment flows. Dependences of
the number-of-nucleon (NN) of the anisotropic flows, $v_1$ and
$v_2$, are surveyed, and the ratio of $v_4$ to $v_2^2$ is studied.

\section{Definition of anisotropic flow and Model Introduction }

\subsection{ Definition of anisotropic flow}

Anisotropic flow is defined as the different $n$-th harmonic
coefficient $v_n$ of an azimuthal Fourier expansion of the
particle invariant distribution \cite{Voloshin}
\begin{equation}
\frac{dN}{d\phi} \propto {1 + 2\sum_{n=1}^\infty v_n cos(n\phi) },
\end{equation}
where $\phi$ is the azimuthal angle between the transverse
momentum of the particle and the reaction plane. Note that in the
coordinate system the $z$-axis along the beam axis, and the impact
parameter axis is labelled as $x$-axis.

The first harmonic coefficient $v_1$ represents the directed flow,
\begin{equation}
v_1 = \langle cos\phi \rangle = \langle \frac{p_x}{p_t} \rangle,
\end{equation}
where $p_t = \sqrt{p_x^2+p_y^2}$ is transverse momentum. $v_2$
represents the elliptic flow which characterizes the eccentricity
of the particle distribution in momentum space,
\begin{equation}
v_2 = \langle cos(2\phi) \rangle = \langle
\frac{p^2_x-p^2_y}{p^2_t} \rangle,
\end{equation}
and $v_4$ represents the 4-th momentum anisotropy,
\begin{equation}
v_{4} =\left\langle \frac{p_{x}^{4}-6p_{x}^{2}p_{y}^{2}+p_{y}^{4}}{%
p_{t}^{4}}\right\rangle . \label{v4}
\end{equation}

In this work, we will study the above flow components.

\subsection{ Nucleonic transport model: QMD Model}

The Quantum Molecular Dynamics (QMD) approach is an n-body theory
to describe heavy ion reactions from intermediate energy to 2 A
GeV. It includes several important parts: the initialization of
the target and the projectile nucleons, the propagation of
nucleons in the effective potential, the collisions between the
nucleons, the Pauli blocking effect and the numerical tests. A
general review about QMD model can be found in \cite{Aichelin}.
The IDQMD model is based on QMD model affiliating the isospin
factors, which includes the mean field, two-body nucleon-nucleon
(NN) collisions and Pauli blocking \cite{Wei,Ma-hbt,Ma3}.

In the QMD model each nucleon is represented by a Gaussian wave
packet with a width $\sqrt{L}$ (here $L$ = 2.16 ${\rm fm}^2$)
centered around the mean position $\vec{r_i}(t)$ and the mean
momentum $\vec{p_i}(t)$,
\begin{equation}
\psi_i(\vec{r},t) = \frac{1}{{(2\pi L)}^{3/4}}
exp[-\frac{{(\vec{r}- \vec{r_i}(t))}^2}{4L}] exp[-\frac{i\vec{r}
\cdot \vec{p_i}(t)}{\hbar}].
\end{equation}

The nucleons interact via nuclear mean field  and nucleon-nucleon
collision. The nuclear mean field can be parameterized by
\begin{equation}
U(\rho,\tau_{z}) = \alpha(\frac{\rho}{\rho_{0}}) +
\beta(\frac{\rho}{\rho_{0}})^{\gamma} +
\frac{1}{2}(1-\tau_{z})V_{c}  + C_{sym} \frac{(\rho_{n} -
\rho_{p})}{\rho_{0}}\tau_{z} + U^{Yuk}
\end{equation}
with $\rho_{0}$ the normal nuclear matter density (0.16 $fm^{-3}$
is used here). $\rho$, $\rho_{n}$ and $\rho_{p}$ are the total,
neutron and proton densities, respectively. $\tau_{z}$ is $z$th
component of the isospin degree of freedom, which equals 1 or -1
for neutrons or protons, respectively. The coefficients $\alpha$,
$\beta$ and $\gamma$ are parameters for nuclear equation of state.    Two set parameters are used:
$\alpha$ = -124 MeV, $\beta$ = 70.5 MeV and $\gamma$ = 2.0 which
corresponds to the so-called hard EOS,
with an incompressibility of $K$ = 380 MeV;
and $\alpha$ = -356 MeV, $\beta$ = 303 MeV and $\gamma$ = 7/6 which
corresponds to the so-called soft EOS with an incompressibility of $K$ = 200 MeV;
$C_{sym}$ is the symmetry energy strength due to the
difference of neutron and proton \cite{Aichelin}, here $C_{sym} = 32$ MeV is used to consider
isospin effects, or $C_{sym} = 0$ for no isospin effect.
$V_{c}$ is
the Coulomb potential,  $U^{Yuk}$ is Yukawa (surface) potential.
In this model, the isospin effects can be included in some terms,
such as in-medium nucleon-nucleon cross section and Pauli blocking \cite{Wei,Ma-hbt}.

The time evolution of the colliding system is given by the
generalized variational principal. Since the QMD can naturally
describe the fluctuation and correlation, we can study the nuclear
clusters in the model \cite{Aichelin,Wei,Ma-hbt,Ma3}.
 In QMD model, nuclear clusters are usually recognized  by a simple
coalescence model: i.e. nucleons are considered to be part of a
cluster if in the end at least one other nucleon is closer than
$\Delta r_{min} \leq 3.5$ fm in coordinate space and $\Delta
p_{min} \leq 300$ MeV/c  in momentum space  \cite{Aichelin}. This
mechanism has been extensively applied in transport theory for the
cluster formation.

\section{Number-of-Nucleon Scaling of the Directed and Elliptic Flows}

Reactions of $^{40}$Ca +$^{40}$Ca, $^{86}$Kr + $^{58}$Ni and
$^{86}$Kr + $^{124}$Sn at 25 MeV/nucleon have been simulated by
IDQMD. To compare the results from different reactions, the
similar centrality has been chosen for the above systems, namely
4-8 fm, 6-10 fm and 7 - 11 fm, respectively. Since the reaction
systems tend to freeze-out around 120 fm/c in the model
calculation, the physics results can be extracted in the
freeze-out stage. In this work, the results are extracted  at 200
fm/c. Some parts of physics results have been reported in our
recent papers ~\cite{Yan,Ma-AIP}.

\begin{figure}
\vspace{-0.6truein}
\includegraphics[scale=0.6]{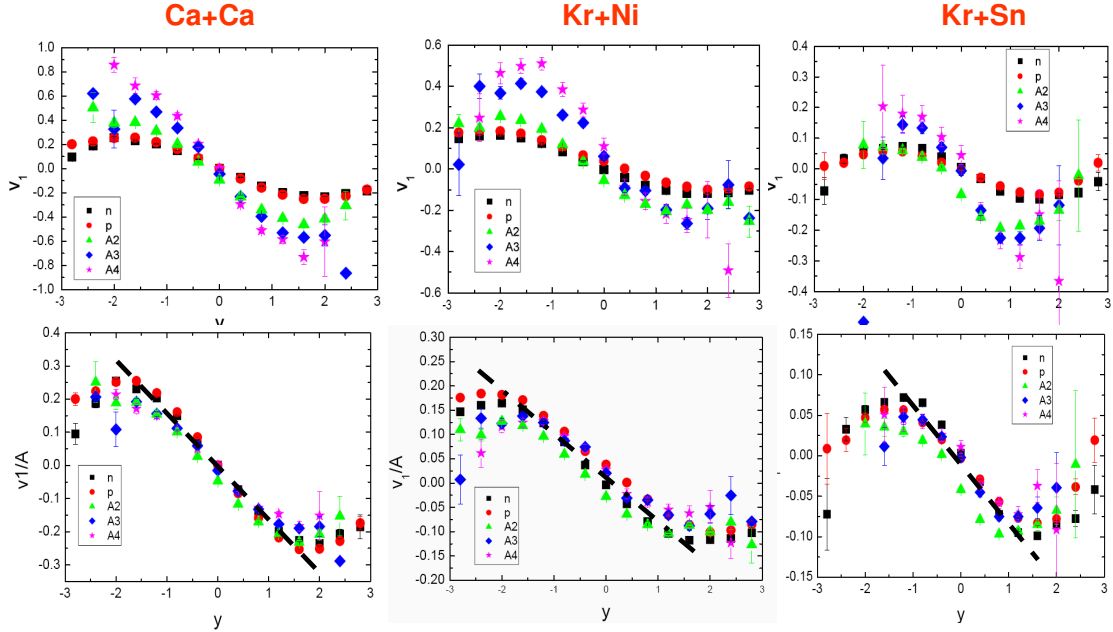}
\vspace{-0.6truein} \caption{\footnotesize Ca+Ca at 4-8 fm (left
panels), Kr + Ni at 6-10 fm (middle panels), Kr + Sn at 7 - 10 fm
(right panels). The upper panels  show $v_1$ versus rapidity and
the lower panels show $v_1$/A, i.e. the number-of-nucleon scaled
directed flow, versus rapidity.  The dashed line guides the eyes.
 }\label{Fig-v1-y}
\end{figure}

The directed flow $v_1$ as a function of rapidity ($y$) for the
above systems has been studied. Fig.~\ref{Fig-v1-y} shows $v_1$
versus rapidity (upper panels) and the nucleon-number scaled
$v_1$/A as a function of rapidity. From this figure, we found that
the slope of the directed flow as a function of rapidity is
negative for three systems at 25 MeV/nucleon, which indicates that
attractive mean filed is important in such a low energy
\cite{yg_prc}. Before the nucleon-number scaling, the values of
directed flow are different for different-mass clusters: the
heavier the clusters, the large the absolute value of directed
flow. However, all the curves of rapidity-dependent $v_1$ almost
collapse onto the same curves by dividing $v_1$ by its number of
nucleon, which illustrates that the directed flow of the light
nuclear clusters satisfies the number-of-nucleon scaling.
Actually, the previous experimental data for light nuclear
clusters up to A = 4 for $^{86}$Kr + $^{197}$Au showed that the
directed flow is approximately proportional to mass number
\cite{Huang}.

\begin{figure}
\vspace{-0.2truein}
\includegraphics[scale=0.6]{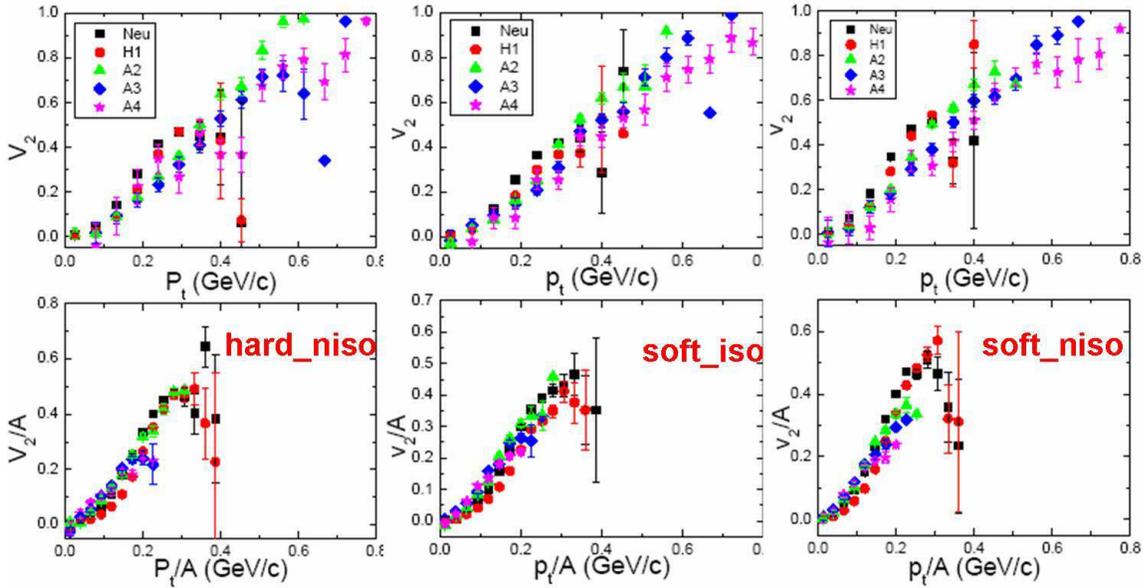}
\vspace{-0.2truein} \caption{\footnotesize The upper panels:
Elliptic flow as a function of transverse momentum ($p_t$) for
$^{86}$Kr + $^{124}$Sn at 25 MeV/nucleon. Three simulations were
done: hard EOS without isospin related terms (hard$_-$niso), soft
EOS with isospin related terms (soft$_-$iso), and soft EOS without
isospin related terms (soft$_-$niso). Squares represent for
neutrons, circles for protons, triangles for fragments of $A$ = 2,
diamonds for $A$ = 3 and stars for $A$ = 4; The lower panels:
Elliptic flow per nucleon as a function of transverse momentum per
nucleon.    } \label{Fig-v2-pt}
\end{figure}

The upper panels of Fig.~\ref{Fig-v2-pt} show transverse momentum
dependence of elliptic flow for mid-rapidity light fragments in
three calculation cases: hard EOS without isospin related terms
(hard$_-$niso), soft EOS with isospin related terms (soft$_-$iso),
and soft EOS without isospin related terms (soft$_-$niso). From
the figure, it shows that elliptic flows are positive and increase
with the increasing of $p_t$ even though the values are a little
different for the different cases. The positive value of $v_2$
reflects that the light clusters are preferentially emitted within
the reaction plane, and particles with higher transverse momentum
tend to be strongly  emitted within in-plane, i.e. stronger
positive elliptic flow. In comparison to the elliptic flow at RHIC
energies, the apparent behavior of elliptic flow versus $p_t$
looks similar, but the mechanism is obviously different. In
intermediate energy domain, collective rotation is one of the main
mechanisms to induce the positive elliptic flow
\cite{yg_prc,Ritter,Peter,Shen,Lacey,He}. In this case, the
elliptic flow is mainly driven by the attractive mean field.
However, the strong pressure which is built in early initial
geometrical almond-type anisotropy due to the overlap zone between
both colliding nuclei in coordinate space will rapidly transforms
into the azimuthal anisotropy in momentum space at RHIC energies
\cite{J.Adams}. In other words, the elliptic flow is mainly driven
by the stronger outward pressure. The lower panels in
Fig.~\ref{Fig-v2-pt} displays the elliptic flow per nucleon as a
function of transverse momentum per nucleon, and it looks that
there exists the number-of-nucleon scaling when $p_t/A < 0.25 $
GeV/$c$. This behavior is apparently similar to the
number-of-constituent-quarks scaling of elliptic flow versus
transverse momentum per constituent quark ($p_t/n$) for mesons and
baryons which was observed at RHIC \cite{J.Adams}.

\section{$v_4/v_2^2$-scaling}
\begin{figure}
\vspace{-.6truein}
\includegraphics[scale=0.6]{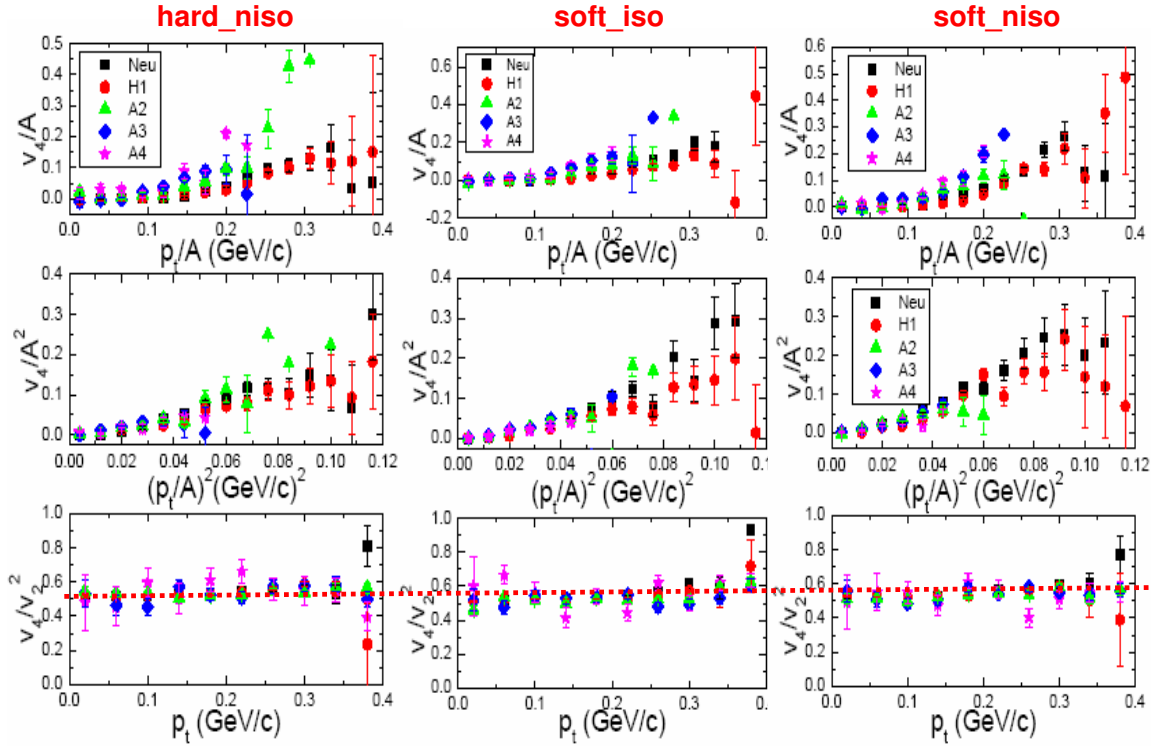}
\vspace{-0.6truein} \caption{\footnotesize The upper panels:
$v_4/A$ as a function of $p_t/A$ for different particles, namely,
neutrons (squares), protons (circles), fragments of $A$ = 2
(triangles), $A$ = 3 (diamonds) and $A$ = 4 (stars) in three
different calculation conditions.  The middle panels:  $v_4/A^2$
as a function of $(p_t/A)^2$.  The lower panels: the ratios of
$v_4/v_2^2$ for different particles vs  $p_t$. Three simulations
were done in this figure as Fig. 2.
 }\label{Fig-v4}
\end{figure}

Recent the RHIC results show that $v_4/v_2^2$ for hadrons keeps an
almost constant which is independent on $p_t$ \cite{STAR-prc2006},
we would like to know what the higher order momentum anisotropy in
the intermediate energy is. In the present work, we explore the
behavior of $v_4$ in the model calculation. Fig.~\ref{Fig-v4}
shows the features of $v_4$ for $^{86}$Kr + $^{124}$Sn at 25
MeV/nucleon for IDQMD simulation with hard$_-$niso, soft$_-$iso
and soft$_-$niso. Similar to the relationship of $v_2/A$ versus
$p_t/A$, we plot $v_4/A$ as a function of $p_t/A$ (the upper
panel). The divergence of the different curves between different
particles in
 $v_4/A$ versus $p_t/A$  indicates no simple scaling of nucleon number
for 4-th momentum anisotropy. However, if we plot $v_4/A^2$ versus
$(p_t/A)^2$ (the middle panels), it looks that the points of
different particles nearly merge together and it means a certain
of scaling holds between two variables. Due to a nearly constant
value of $v_4/v_2^2$ in the studied $p_t$ range (see the bottom
panels in Fig.~\ref{Fig-v4}) together with the number-of-nucleon
scaling behavior of $v_2/A$ vs $p_t/A$, $v_4/A^2$ should scale
with $(p_t/A)^2$, as shown in the middle panels of
Fig.~\ref{Fig-v4}. Interestingly, the constant value of
$v_4/v_2^2$ does not depend on the EOS and isospin dependent term,
which may indicate this constant value is robust for intermediate
energy heavy ion collision.

The RHIC experimental data demonstrated a scaling relationship
between 2-nd flow ($v_2$) and n-th flow ($v_n$), namely
$v_{n}(p_{t})\sim v_{2}^{n/2}(p_{t})$ \cite{STAR03}. It has been
shown \cite{Kolb2,ChenLW} that such  scaling relation follows from
a naive quark coalescence model \cite{Molnari} that only allows
quarks with equal momentum to form a hadron. Denoting the meson
anisotropic flows by $v_{n,M}(p_t)$ and baryon anisotropic flows
by $v_{n,B}(p_t)$, it was found that $v_{4,M}(p_{t}) =
(1/4)v_{2,M}^{2}(p_{t})$ for mesons and $v_{4,B}(p_{t}) =
(1/3)v_{2,B}^{2}(p_{t})$ for baryons if quarks have no
higher-order anisotropic flows. Since mesons dominate the yield of
charged particles in RHIC  data, the smaller scaling
factor of $1/4$ than the empirical value of about $1$ indicates
that higher-order quark anisotropic flows cannot be neglected.
Including the latter contribution, one can show that
\begin{equation}
\frac{v_{4,M}}{v_{2,M}^{2}} \approx \frac{1}{4}+\frac{1}{2}%
\frac{v_{4,q}}{v_{2,q}^{2}},\allowbreak \label{v4Mscal} ~~~~
\frac{v_{4,B}}{ v_{2,B}^2} \approx \frac{1}{3} + \frac{1}{3}
\frac{v_{4,q}}{v_{2,q}^2}\,, \label{eqn-v4}
\end{equation}
where $v_{n,q}$ denotes the quark anisotropic flows. The meson
anisotropic flows thus satisfy the scaling relations if the quark
anisotropic flows also satisfy such relations.
One can go one step
further and assume that the observed scaling of the hadronic $v_2$
actually results from a similar scaling occurring at the partonic
level. In this case, if one assumes \cite{Kolb2,ChenLW} that the
scaling relation for the partons is as follows:
\begin{equation}
v_4^q = (v_2^q)^2, \label{eq-parton}
 \end{equation}
 and then hadronic ratio
$v_4/v_2^2$ then equals $1/4 + 1/2 = 3/4$ for mesons and
$1/3+1/3=2/3$ for baryons, respectively.

If we assume the scaling laws of mesons (NCQ=2)
and baryons (NCQ=3) (Eq.~\ref{eqn-v4}) are also valid for A = 2 and 3 nuclear
clusters, respectively, then $v_4/v_2^2$ for A = 2 and 3 clusters
indeed give the same value of 1/2 as nucleons, as shown in
Fig.~\ref{Fig-v4}(c). Coincidentally the predicted value of the
ratio of $v_4/v_2^2$ for hadrons is also 1/2 if the matter
produced in ultra-relativistic heavy ion collisions reaches to
thermal equilibrium and its subsequent evolution follows the laws
of ideal fluid dynamics \cite{Bro}. It is interesting to note the
same ratio was predicted in two different models at very different
energies, which is of course worth to be further investigated in
near future.

\section{Summary}

In summary, we have investigated the behavior of anisotropic
flows, namely $v_1$, $v_2$ and $v_4$,  for light nuclear fragments
produced by $^{40}$Ca +$^{40}$Ca, $^{86}$Kr + $^{58}$Ni and
$^{86}$Kr + $^{124}$Sn at 25 MeV/nucleon for peripheral collisions
in the framework of quantum molecular dynamics model. $v_1$ shows
a negative slope versus  rapidity, indicating the attractive mean
field plays a dominant role in directed flow. Both $v_2$ and $v_4$
generally show positive values and increase with $p_t$. The
positive $v_2$ illustrates that the in-plane emission is
preferential. By the scaling of the number-of-nucleons, both
$v_1/A$ versus rapidity and $v_2/A$ versus $p_t/A$ for light
nuclear fragments approximately collapse on the similar curve,
respectively, which means that there exists directed flow and
elliptic flow scalings on the number-of-nucleon. Nucleon-number
scaling originates from nucleonic coalescence. For 4-th momentum
anisotropy $v_4$, it seems to be scaled by $v_2^2$, and
$v_4/v_2^2$ is a constant, i.e. $~$1/2, for light nuclear
fragments. Comparing our above predictions with intermediate
energy HIC data for the  flow scalings which is presented in this
work is expected to shed light on nucleonic collectivity and its
rich features.

\section{ Acknowledgements}
This work was supported in part by the Shanghai Development
Foundation for Science and Technology under Grant Numbers
05XD14021 and 06JC14082, the National Natural Science Foundation
of China under Grant No 10535010, 10328259 and 10135030.

\end{document}